\documentclass[useAMS,usenatbib,usegraphicx]{mn2e}  
\usepackage[usenames]{color}

\newcommand{\be}{\begin{equation}}
\newcommand{\ee}{\end{equation}}
\newcommand{\bea}{\begin{eqnarray}}
\newcommand{\eea}{\end{eqnarray}}

\title[DM in MACS0717] 
{Hubble Frontier Field Free-Form Mass Mapping of the Massive Multiple-Merging Cluster MACSJ0717.5+3745} 
\author[J.M. Diego]  
  {Jose M. Diego\footnote{jdiego@ifca.unican.es}$^1$, Tom Broadhurst$^{2,3}$, Adi Zitrin$^{4,5}$,Daniel Lam$^6$,Jeremy Lim$^6$, 
  \newauthor
  Holland C. Ford$^7$, Wei Zheng$^7$\\
$^{1}$IFCA, Instituto de F\'isica de Cantabria (UC-CSIC), Av. de Los Castros s/n, 39005 Santander, Spain\\
$^{2}$Fisika Teorikoa, Zientzia eta Teknologia Fakultatea, Euskal Herriko 
Unibertsitatea UPV/EHU\\ 
$^3$IKERBASQUE, Basque Foundation for Science, Alameda Urquijo, 36-5 48008 Bilbao, Spain\\
$^4$ Cahill Center for Astronomy and Astrophysics, California Institute of Technology, MS 249-17, Pasadena, CA 91125, USA.\\
$^5$ Hubble Fellow\\
$^6$ Department of Physics, The University of Hong Kong, Pokfulam Road, Hong Kong \\
$^7$ Dept. of Physics and Astronomy, Johns Hopkins University, Baltimore, Maryland, USA
}

\date{Draft version \today}  
  
\pagerange{\pageref{firstpage}--\pageref{lastpage}}  
  
\begin{document}  
\maketitle  
 
\label{firstpage}  
\begin{abstract} 
We examine the latest data on the cluster MACSJ0717.5+3745 from the Hubble Frontier Fields campaign. The critically lensed area is the largest known of any lens and very irregular making it a challenge for parametric modelling. Using our Free-Form method we obtain an accurate solution, identify here many new sets of multiple images, doubling the number of constraints and improving the reconstruction of the dark matter distribution. Our reconstructed mass map shows several distinct central substructures with shallow density profiles, clarifying earlier work and defining well the relation between the dark matter distribution and the luminous and X-ray peaks within the critically lensed region.
Using our free-form method, we are able to meaningfully subtract the mass contribution from cluster members to the deflection field to trace the smoothly distributed cluster dark matter distribution. We find 4 distinct concentrations, 3 of which are  coincident with  the luminous matter. The fourth peak has a significant offset from both the closest luminous and X-ray peaks. These findings, together with dynamical data from the motions of galaxies and gas  will be important for  uncovering the potentially important implications of this extremely massive and intriguing system. 

\end{abstract}  
  
\section{Introduction}\label{sect_intro}  

The standard model of structure formation is built on the conclusion that about 85\% of the mass in the universe 
is of an unknown form which only gravitates. 
The standard interpretation of this dark matter as massive fermionic particles has, so far, no experimental evidence from sensitive 
direct searches via nuclear recoil  \citep{Akerib2014} nor have such particles been generated at high energies 
with the LHC. Astronomically, the most extreme effects of dark matter can be found in massive galaxy clusters, where the general relativistic warp-ing of 
spacetime leads to extreme lensing distortions on a scale far in excess of 
that due to the observed stellar  or gaseous cluster material. Among all known clusters, MACSJ0717.5+3745 (MACS0717 hereafter, \cite{Ebeling2007}) 
is one of the most massive and extreme clusters in terms of its mass and temperature, with light deflections of over 
an arcminute discovered by \cite{Zitrin2009} (Z09 hereafter). This cluster has been exten-sively studied from a multiwavelength perspective providing a 
unique opportunity to explore the interplay between the visible and dark matter.

\begin{figure*}    
 {\includegraphics[width=18.0cm]{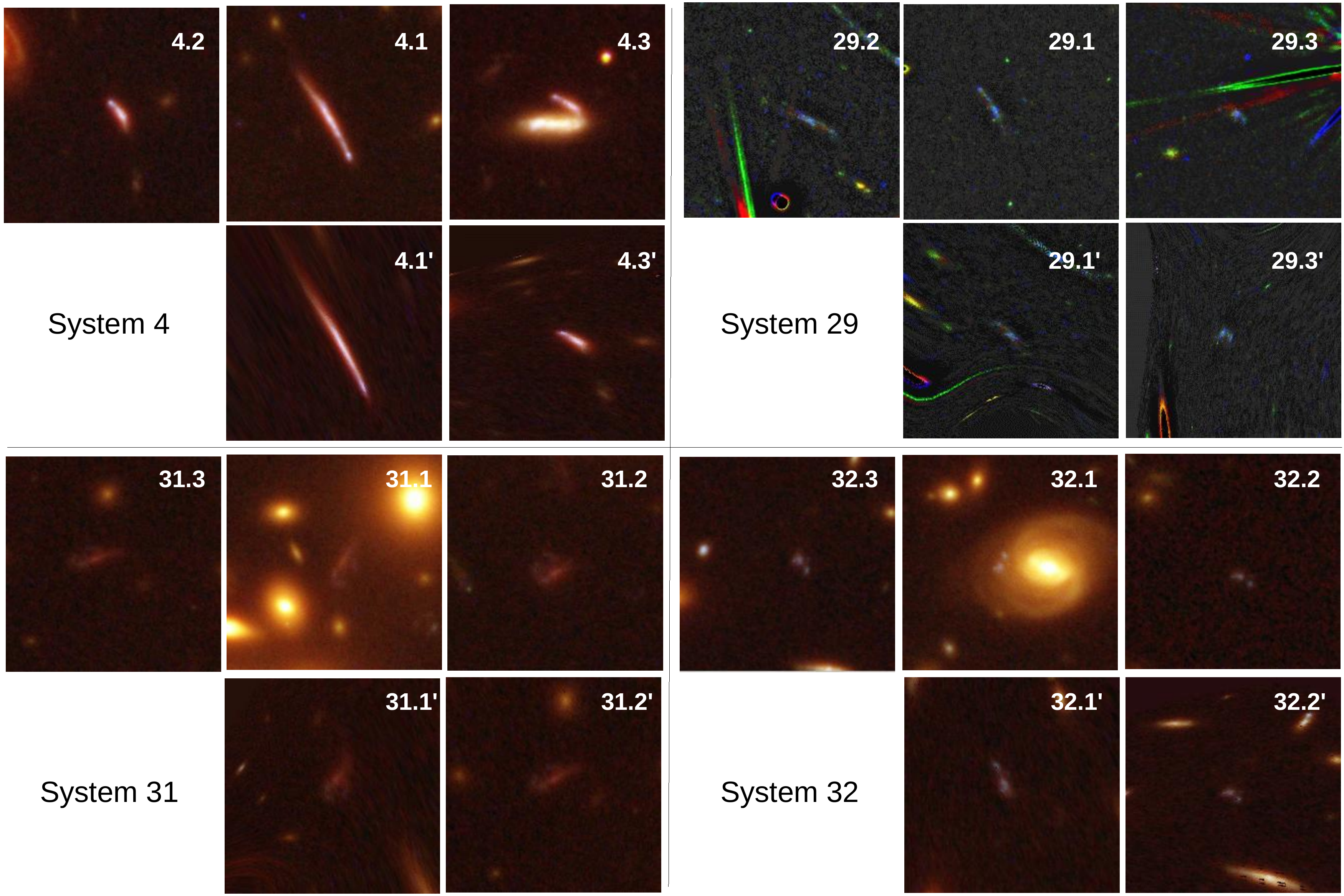}}
   \caption{Examples of lensed systems compared with the prediction from our model. 
            We show 4 systems.
            For each case, the top row contains three counter-images, with the first one on the left being used 
            to predict the lensed counter-images (bottom) for which the predicted details are readily recognised in these cases. Each stamp corresponds to 6.6 arcseconds across. }
   \label{fig_lensed_systems}  
\end{figure*}  

Radio observations reveal that this cluster hosts one the most powerful radio halos known to date \citep{vanWeeren2009,Bonafede2009,Pandey2013}. 
At microwave wavelengths, this cluster is a bright Sunyaev-Zeldovich effect source \citep{Mroczkowski2012} which has enabled an estimate of 
its radial velocity through the Doppler shift induced by the plasma in the cluster to the photons of the cosmic microwave background \citep{Sayers2013}. 
In the optical, a filament of galaxies seems to extend over cosmic scales from the centre of this cluster \citep{Ebeling2004,Medezinski2013}. 
The same filament can be observed in weak lensing maps  \citep{Jauzac2012,Medezinski2013} confirming that this cluster represents a "node" 
of the cosmic web. Strong gravitational lensing has has uncovered a complex structure where the distribution of 
lensed images is more anatomical in shape than it is geometrical ( Zitrin etal 2008) so that the strong lensing was not recognised for years despite adequate archival data.  
Several subclusters  may by in the process of  simultaneously converging to form one of the largest, complex and most extreme clusters in the 
universe \citep{Zitrin2009,Jauzac2012,Limousin2012,Mroczkowski2012}.  According to \cite{Zitrin2009}, this cluster is the largest lens known to date, with an effective Einstein 
radius of $\approx 55$ arcseconds.  The extreme nature of this clusters is better appreciated in X-rays where the plasma temperature in places may exceed, 
20 keV \citep{Ma2009}. The high temperature regions observed in X-rays correlate well with with shocks that are detected in radio maps \citep{Bonafede2009,Pandey2013}. 
Radio observations have also confirmed that the radio emission is polarized, indicating that the magnetic field is ordered on large scales. The alignment of the radio 
halo perpendicular to the long axis of the dark matter distribution suggests that the radio emission is the result of a merger-related shock wave, with the emitting 
particles being shock accelerated  \citep{vanWeeren2009}.  The SZ effect is well mapped in this central region with high resolution Mustang data \citep{Mroczkowski2012}   
confirming the pressure enhanced shocked gas and they find a high line of sight velocity for a central gas component of +3600km/s, 
from a claimed detection of the Kinetic SZE effect, in agreement with the internal galaxy velocity analysis of \cite{Ma2009},   
and nearly orthogonal to the long axis of the dark matter, implying multiple merging.

In this paper we use the recently released data from the Hubble Frontier Fields program\footnote{http://www.stsci.edu/hst/campaigns/frontier-fields/}  
(or HFF hereafter) on this cluster. We use the new data to identify new multiply-imaged systems that are later used to constrain the mass model of this cluster with an unprecedented number of constraints. We use our robust free-form method that does not rely on major assumptions about the distribution of dark matter other than the safe assumption 
that the galaxy members contain some mass and that more luminous galaxies correspond in general to more massive galaxies.

\begin{figure}  
   \includegraphics[width=8cm]{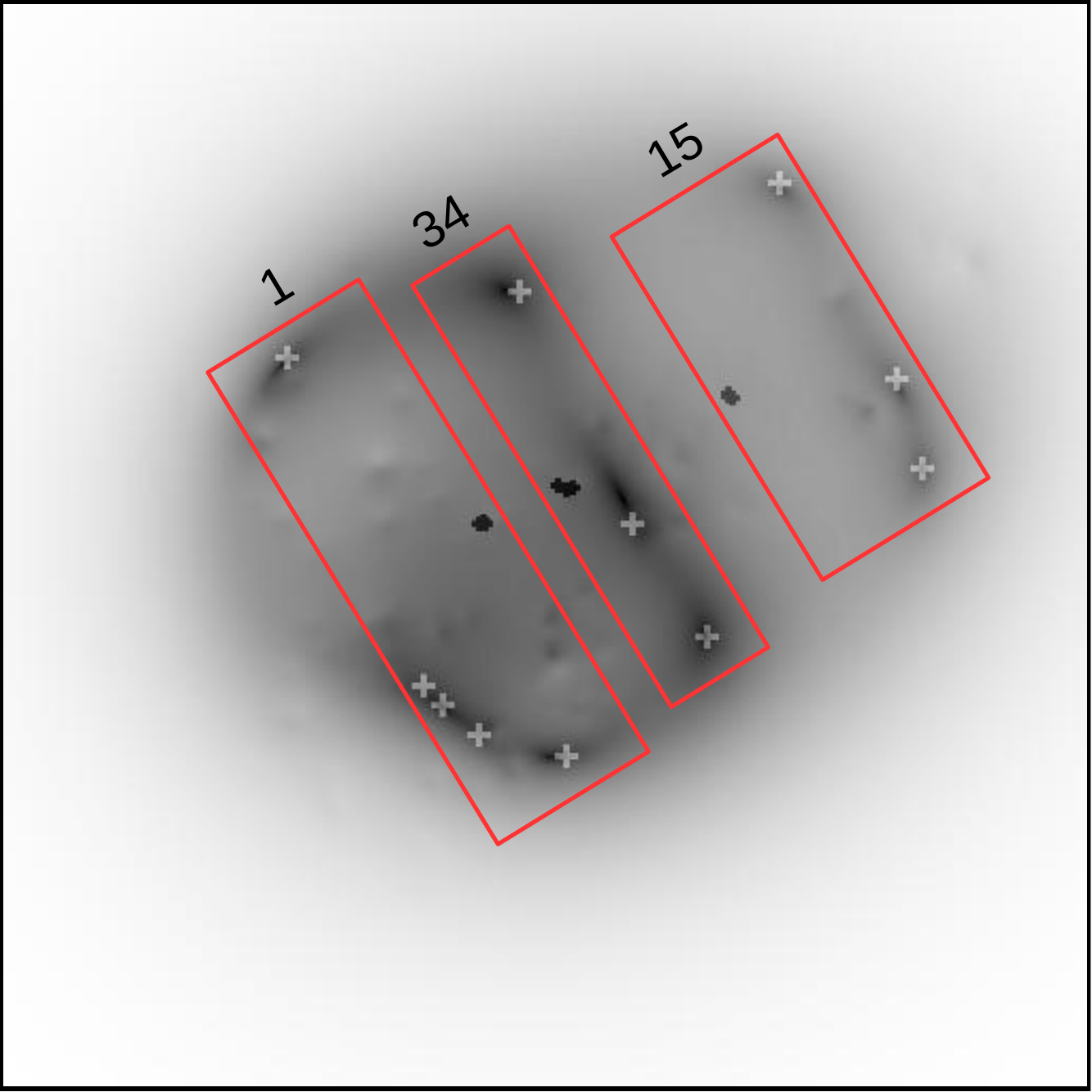} 
   \caption{Examples of model predictions for systems 1, 15 and 34. Light grey crosses mark the position of the observed arcs. Darker regions mark the 
            predicted position in the image plane of the arcs from our lens model. Dark crosses mark the predicted position of the observed arcs in the source plane.}
    \label{fig_predictions}  
\end{figure}  

The paper is organized as follows. We describe the Hubble data in section \ref{sect_S2}.
The lensing data is described in section \ref{sect_S3}.
In section \ref{sect_S4} we give a brief description of the reconstruction method.
Section \ref{sect_S5} presents the results of the lensing analysis.
The robusteness of our solution is discussed in more detail in section \ref{sect_S6}.
We discuss our results in section \ref{sect_S7}, and finally we conclude in section \ref{sect_S8}.

Throughout the paper we assume a cosmological model with $\Omega_M=0.3$,
$\Lambda=0.7$, $h=70 km/s/Mpc$. For this model, 1 arcscec
equals $6.46$ kpc at the distance of the cluster.

\section{HFF data}\label{sect_S2}

In this paper we used public imaging data obtained from the ACS (filters: F435W, F606W and F814W) 
and the WFC3 (F105W, F125W, F140W and F160W), retrieved from the Mikulski Archive for Space 
Telescope (MAST). The data used in this paper consists of $\approx 1/3$ of the data to be collected. 
Part of the data comes from CLASH \citep{Postman2012}. This release includes the first 40 orbits of
observations of MACS0717 from the Frontier Fields program ID 13498
(PI: J. Lotz), also including archival ACS and WFC3/IR data (programs
9722 and 10420, PI.: H. Ebeling; programs 10493 and 10793, PI.: A.
Gal-Yam; program 12103, PI.: M. Postman; program 13389, PI.: B.Siana;
and program 13459, PI.: T. Treu). Previous observations from earlier shallow 
imaging (14 orbits in total) are also included in the release , together with the 40 orbits from the HFF campaign.
The current release contains 15 orbits in the band F435W, 1 orbit in the band F606W,  
30 orbits in the band F814W and 8 orbits in the WFC3 infrared (IR) bands. The relatively low number of orbits 
in the IR bands makes the current release not ideal for detecting high redshift objects but the deep 
band at F814W allows us to reliably detect intermediate redshift candidates. In the IR bands we use the background corrected 
images, corrected for a time-dependent increase in the background sky level (see for instance \cite{Koekemoer2013}). 
In the optical bands we use the self-calibrated images with improved low-level noise. 
 
From the original files, we produce two sets of color images combining the optical and IR bands. The first set is based on the raw data 
while in the second set we apply a low-pass filter to reduce the diffuse emission form member galaxies. The second set 
is particularly useful to match colors in objects that lie behind a luminous member galaxy.

\begin{figure*}  
 \centerline{ \includegraphics[width=16cm]{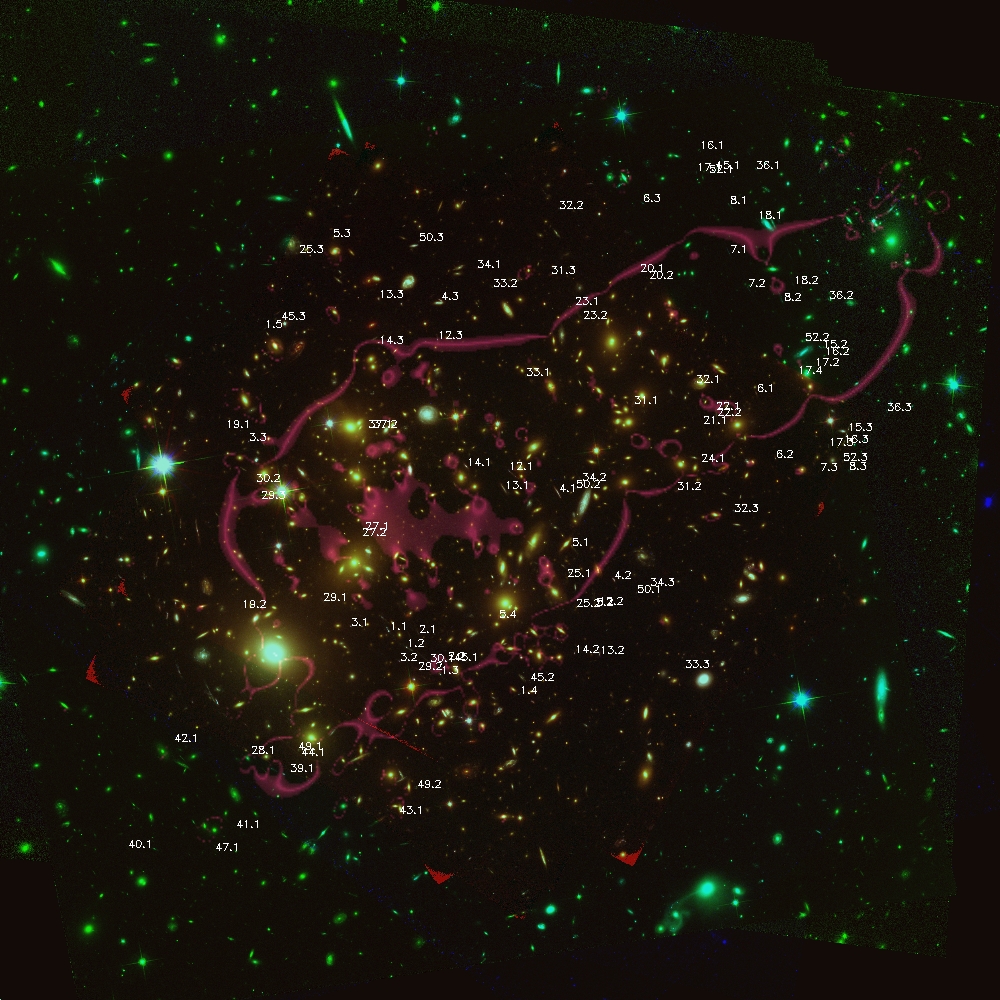}}  
   \caption{Compilation of systems used in the reconstruction and presented in the appendix. 
            The critical curve ($z_s=3$) for model iV desxcribed in section \ref{subsect_models} is shown superimposed in fuschia ($z_s=3$) 
            The field of view is 4x4 arcminutes$^2$ and north is up.} 
   \label{fig_SystemsI}  
\end{figure*}

\section{Lensing data}\label{sect_S3}
Our starting set of strong lensing data is primarily based on the system identification of \cite{Zitrin2009} (Z09). 
Some systems in Z09 are updated with new spectroscopic redshifts from \cite{Limousin2012,Schmidt2014} and \cite{Vanzella2014}, and also 
some system definitions are improved following \cite{Limousin2012}, \cite{Medezinski2013} and \cite{Richard2014}. 
System 2 was excluded in some previous analyses since this source was too faint. The new and much deeper HFF data confirms this 
system as a reliable one so we include it in our analysis.  
Using this set of systems we build a lens-model that is later used to find new candidates taking advantage 
of the deeper Hubble observations. Many system candidates can be found in the new data set. 
In this paper, however, we rely only on the most robust sub-sample. This robust sub-sample is defined after we 
require that the system candidates must have similar colors and morphological features. Also, these systems 
must be consistent in terms of location in the lens plane and parity with the lens model derived in our first step. 
In the process of identifying new candidates we need to assume a redshift for the systems. The lower numebr of orbits of the IR bands 
and the V band (F606W) compared with the F814W band does not allow for precise photometric redshifts of faint objects. 
Redshifts predicted by the lens models have demonstrated its 
usefulness and can be competitive with photometric redshifts as shown in \cite{Lam2014}. This is particularly true for clusters with shallow mass profiles 
(like MACS0717) where the shallow mass distribution makes the location of image pairs extraordinarily sensitive to their redshift.
We identify 17\footnote{At the time of submission of this paper new optical data released through the MAST archive has allowed 
us to uncover 13 system candidates in addition to the 17 new systems used in this work. 
These 13 new systems have not been used in this paper but are included in table A2 in the appendix for completeness. 
Stamps of these new systems are also provided in the website http://www.ifca.unican.es/users/jdiego/MACS0717.} 
new multiple systems that roughly double the original 16 systems in \cite{Zitrin2009,Limousin2012,Richard2014}. 
In addition to the new multiple systems we include also 10 elongated arclets 
(with no identified counter-image) that are helpful to constrain the mass distribution around the critical curves and beyond the Einstein radius. 
The inclusion of the elongated arclets in our lens model are useful since they incorporate important information about the magnification. This is of 
particular interest in the regions beyond the Einstein radius where the lensing constraints from multiply lensed images disappear. 
We include the recently confirmed (spectroscopically) system at z=6.387 as system 19 (following \cite{Richard2014} notation). 
Our model makes a clear prediction for a third image for this system but it could not be found with the current data. 
Also, the magnification for the third image is predicted to be significantly smaller ($\mu \approx 2$) than for the other two images. 
We should note that in previous works, the original system 19 was considered by other authors 
as part of our system 18 at z=2.4 but the new system 19 at z=6.387 is completely different (and already used in \cite{Richard2014}). 
Also, a third counterimage 19.3 is proposed in \cite{Richard2014}, very closed to our predicted position. 
Howevere several faint sources can be seen in that area. The HFF can not confirm nor reject this candidate as it 
is very faint. Hence we do not use the candidate 19.3 from \cite{Richard2014} in our analysis. 
More arclets can be identified in the new Hubble images that will be incorporated in future works together with some of the candidates not used in 
this work and that are expected to be confirmed with the future Hubble data. Our complete strong lensing data set is listed in table A1 in the appendix. 
Color stamps of the full data set can be found in this website\footnote{http://www.ifca.unican.es/users/jdiego/MACS0717}. Additional useful material 
is also included in the same website. 

\begin{figure}  
   \includegraphics[width=8cm]{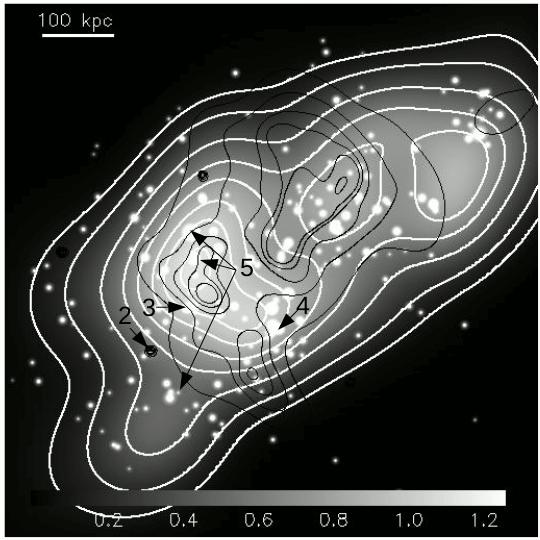} 
   \caption{Reconstructed mass in MACSJ0717 in units of its convergence, $\kappa$, for $z_s=3$. This solution corresponds to case IV described 
            above. The other cases look very similar. 
            The convergence maps have been saturated beyond k=1.25 for clarity purposes. 
            The white contours corresponds to $\kappa=0.1,0.2,0.4,0.6,0.75,0.9,1.0,1.1,1.15$ respectively. The last level overlaps with the saturation level 
            so it can not be appreciated in this plot.
            The black contours represent the X-ray data from Chandra. X-rays have been smoothed with {\small ASMOOTH} \citep{Ebeling2006}.
            The numbers show the galaxies which belong to each layer (2 to 5). The galaxies that are not marked belong to layer 1.} 
    \label{fig_Mass_contour}  
\end{figure}

A few examples\footnote{Additional examples can be found in this website http://www.ifca.unican.es/users/jdiego/MACS0717} 
showing re-lensed images involving the previously known system 4 and the new systems 29, 31 and 32 are shown in 
figure \ref{fig_lensed_systems}. In general, we find a very good agreement between the re-lensed systems and the observed ones. In some cases, 
differences form the observed arcs highlight deficiencies in our lens model. For instance, the relensed system 32.1' in figure  \ref{fig_lensed_systems} differs 
in orientation and magnification from the observed one (32.1). In this case, the spiral galaxy shown in the stamp 32.1 is not included in our lens model that only 
includes elliptical galaxies. As shown by \cite{Diego2014c}, this system could be used effectively to constrain the mass profile of the spiral galaxy, which 
is acting as a secondary lens.  

The relative positions of the images defining our lensing data set is shown in figure \ref{fig_SystemsI}.
Figure \ref{fig_predictions} shows examples of model predictions for three systems. The grey regions represent 
the distances (in the source plane but translated into the image plane) between the model prediction and the position of the source (this position is also 
part of our solution, or array $X$, described below in section \ref{sect_S4}). The darker the region the more likely a lensed image is to be observed at that position. 
The positions of the observed arcs are marked with light grey crosses while the position of the observed arcs projected back into the source plane are marked with 
dark crosses. Based on our model predictions, color information, morpholohy of the galaxies and model-basd parity, we are able to identify all the additional 
multiple images listed in the appendix. The model predictions are useful to identify possible problesm with teh system identofications like in system 34 in  
figure \ref{fig_predictions}. The position of the arcs in the source plane do not agree as well as in the other systems higlighting a possible problem with our lens model in 
this particular region of the cluster or that the redhsift of this system may need to be revised.  The new HFF data will help clarify some of these systems and also allow 
for a firmer identification of other system candidates not used in this work. The model predictions for all the systems can be found in the 
website\footnote{Additional examples can be found in this website http://www.ifca.unican.es/users/jdiego/MACS0717} with the support material. These model predictions 
show also the expected position of the counterimages that could not be identified wity the current data. Some of these counterimages are predicted to be demagnified versions 
of the background galaxies buried in the cluster members and hence very unlikely to be uncovered. 

Figure \ref{fig_SystemsI} shows the full data set together with the critical curve for one of our models (case IV 
described below in section \ref{subsect_models}).  

\begin{figure*}  
   \includegraphics[width=15cm]{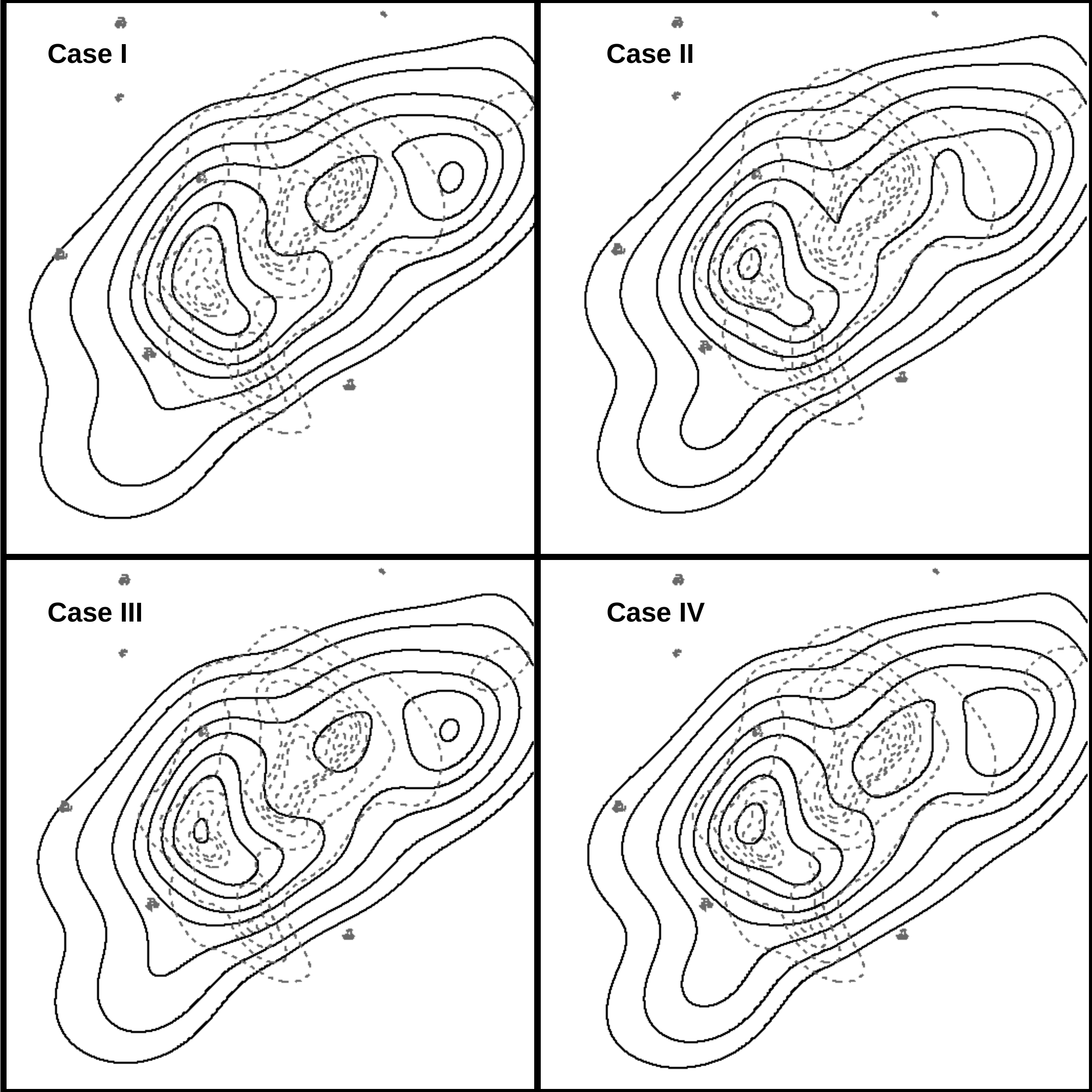} 
   \caption{Like in figure \ref{fig_Mass_contour} but comparing the dark matter to the X-ray contours for the 4 different solutions.
           The DM component corresponds to the total mass map minus the galaxy component. The compact grey sources ar X-ray sources (see text). } 
    \label{fig_Mass_contour2}  
\end{figure*}  

Although we are confident about our new system identification (they all are consistent with the lens model and show similar colors), 
it its possible that some corrections are made when the future data arrives (we should remind that we are using 1/3 of the planned data), 
specially for the unresolved galaxies for which the lack of morphological information does not allow for a more firm confirmation. 
We should however highlight that since these systems are consistent with the lens model, even in the situation where a few systems are wrongly 
matched,  we don't expect significant changes in or lens model and/or conclusions in this work.

\section{Reconstruction method}\label{sect_S4}
We use the improved method, WSLAP+, to perform the mass reconstruction. 
The reader can find the details of the method in our previous papers
\citep{Diego05a,Diego05b,Diego2007,Ponente2011,Sendra2014,Lam2014,Diego2014a,Diego2014b}. 
Here we give a brief summary of the most essential elements. \\

Given the standard lens equation, 
\begin{equation} \beta = \theta -
\alpha(\theta,\Sigma), 
\label{eq_lens} 
\end{equation} 
where $\theta$ is the observed position of the source, $\alpha$ is the
deflection angle, $\Sigma(\theta)$ is the surface mass density of the
cluster at the position $\theta$, and $\beta$ is the position of
the background source. Both the strong lensing and weak lensing
observables can be expressed in terms of derivatives of the lensing
potential. 
\begin{equation}
\label{2-dim_potential} 
\psi(\theta) = \frac{4 G D_{l}D_{ls}}{c^2 D_{s}} \int d^2\theta'
\Sigma(\theta')ln(|\theta - \theta'|), \label{eq_psi} 
\end{equation}

where $D_l$, $D_s$, and $D_{ls}$ are the
angular diameter distances to the lens, to the source and from the lens to 
the source, respectively. The unknowns of the lensing
problem are in general the surface mass density and the positions of
the background sources. As shown in \cite{Diego05a}, the
strong lensing problem can be expressed as a system of linear
equations that can be represented in a compact form, 
\begin{equation}
\Theta = \Gamma X, 
\label{eq_lens_system} 
\end{equation} 
where the measured strong lensing observables are contained in the
array $\Theta$ of dimension $N_{\Theta }=2N_{SL}$, the
unknown surface mass density and source positions are in the array $X$
of dimension $N_X=N_c + N_g + 2N_s$ and the matrix $\Gamma$ is known
(for a given grid configuration and fiducial galaxy deflection field) 
and has dimension $N_{\Theta }\times N_X$.  $N_{SL}$ is the number
of strong lensing observables (each one contributing with two constraints,
$x$, and $y$), $N_c$ is the number of grid points (or cells) that we use to divide
the field of view. In this paper we consider a regular grid of $N_c=32\times32=1024$ cells 
covering the field of view shown in figure \ref{fig_SystemsI} (4 arcminutes).
Each grid point contains a Gaussian function. The width of the Gaussians are chosen in such a way 
that two neighbouring grid points with the same amplitude produce a horizontal plateau in between the two 
overlapping Gaussians.
$N_g$ is the number of deflection fields (from cluster members) that we consider.  
In this work we test two different configurations for 
the deflection field where  $N_g$ is equal to 2 (all member galaxies conform a unique 
deflection field except one foreground galaxy that is placed in a different redshift) or $N_g=5$ which corresponds to the case 
where some bright galaxies are treated in an independent way from the rest of the cluster members. Details of these two 
configurations are given in the next subsection.
$N_s$ is the number of background sources (each contributes with two unknowns, 
$\beta_x$, and $\beta_y$). The solution is found after
minimising a quadratic function that estimates the solution of the
system of equations \ref{eq_lens_system}.  For this minimisation we
use a quadratic algorithm which is optimised for solutions with the
constraint that the solution, $X$, must be positive. Since the vector $X$ contains the grid masses, 
the re-normalisation factors for the galaxy deflection field and the background source positions, and all these 
quantities are always positive (the zero of the source positions is defined in the bottom left corner of the 
field of view), imposing  $X>0$ helps in constraining the space of meaningful solutions. 
The condition $X>0$ also helps in regularising the solution as it avoids large negative and positive 
contiguous fluctuations. 

\subsection{Member galaxy deflections}\label{subsect_models}

The member galaxies defining our fiducial galaxy field are all elliptical galaxies selected from the red sequence.  
The most luminous galaxy in the Hubble image is a known foreground galaxy (2MASX J07173724+3744224) at z=0.1546 \citep{Pandey2013}. 
This galaxy, although well in the foreground  ii is still relatively Luminous and hence its mass may have a non-negligible effect in the lens model. Following \cite{Zitrin2009} we include this galaxy in our lens model but set this galaxy to be at z=0.1546. The remaining galaxies in the cluster are assigned a fiducial mass based on their flux. Given the galaxy member positions and masses, $M_i$, we assume a NFW profile for each galaxy with a scale radius proportional to $M_i^{1/3}$. 
A fiducial deflection field is then computed summed from these galaxy members, as their deflections add linearly via the lensing equation. The galaxies are "split" into different layers to account 
for possible projection effects and/or some variaiton in the galaxy luminosity-to-mass ratio. Our method re-scales this fiducial field by a constant C per layer and combines it with the fiducial field from a gridded mass distribution to reproduce the observed positions and magnifications of the multiply 
lensed systems and arclets. For clarity, the layers used in this work are explicitly shown in figure \ref{fig_Mass_contour} and these additional normalisation parameters are accounted for when calculating the chi-square  fit to the data.

The selected galaxies are shown in figure \ref{fig_Mass_contour}. In the same plot we also indicate the galaxies that conform the 
different layers. We explore 4 different solutions according to the number of layers used in the reconstruction and/or the number of iterations 
in the minimization process. \\

I) This is the simplest model with two layers for the fiducial field of galaxies. 
   Layer one contains all the galaxy cluster members at z=0.546 and layer 2 contains the foreground galaxy at z=0.1546. 
   The solution is obtained after 5000 iterations of the code. This range of iterations was proven in earlier works to be a safe 
   number to avoid overfitting.\\
II) as in case I above but we double the number of iterations to 10000. In most cases, at this regime signs of overfitting start to be evident in 
    the solution like oversized radial critical curves. However, at 10000 iterations these signs are still not present in the solution. \\\
III) as in case I (5000 iterations) but considering 5 layers instead of 2. Like in case I layer 2 contains the foreground galaxy at z=0.1546, 
     layer 3 contains a big elliptical galaxy at its centre, layer 4 contains another large galaxy south-west of the central galaxy, layer 
     5 contains 3 large galaxies, 2 above and one below the central galaxy. Finally layer 1 contains all the remaining galaxies.   \\
IV) as in case III but for 10000 iterations.

\begin{figure}  
   \includegraphics[width=8cm]{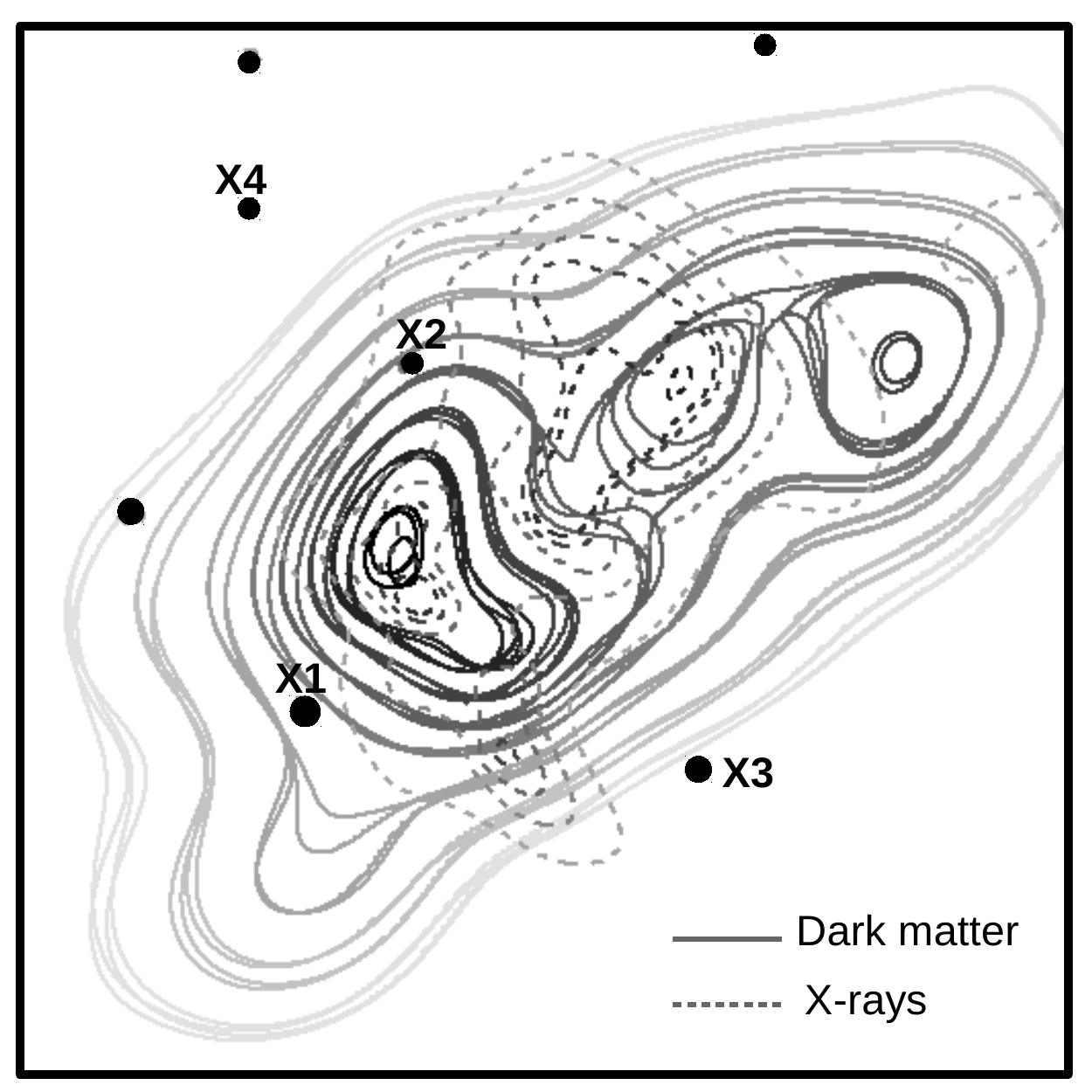} 
   \caption{Like in figure \ref{fig_Mass_contour2} but comparing the 4 solutions for the DM distribution 
            (solid contours) with the X-ray distribution from Chandra (dashed contours)  shown for comparison. The points X1,X2,X3, and X4 
            mark the position of some bright X-ray sources (see text).}
    \label{fig_Mass_contour_X}  
\end{figure}

\section{Dark matter distribution}\label{sect_S5}

The mass reconstruction is shown in figure \ref{fig_Mass_contour} for the solution obtained in case IV discussed in the previous section.
Together with the mass we show the X-ray emission from Chandra. The Chandra data has been smoothed using the widely used
code {\small ASMOOTH}, \citep{Ebeling2006}. The solutions for the other 3 cases are shown in figure \ref{fig_Mass_contour2}.  
The contours in figures \ref{fig_Mass_contour}, and \ref{fig_Mass_contour2} correspond to the {\it smooth} component 
of the mass distribution obtained after subtracting the galaxy contribution to the mass map from the total mass. 
The first conclusion we can derive from these results is that the mass distribution shows multiple distinct concentrations as shown in Figures \ref{fig_Mass_contour} and \ref{fig_Mass_contour2}. 
Earlier work has generally claimed 4 peaks in this region \citep{Ma2009,Limousin2012,Medezinski2013}. 
A more precise picture of the location of our peaks  can be seen in figure \ref{fig_Deconvolved} below. 
Three of  our detected maxima correlate well with the observed galaxy and gas enhancements as seen in Figure \ref{fig_Deconvolved}. The third peak on the right side of the image (west) shows 
a significant offset both with the galaxies and X-rays. Some caution should be exercised with regard to this third peak as we observe have no lensing constraints to help beyond this peak.  
This  subgroup of galaxies does coincide with an X-ray emission peak whereas our mass peak does not , so this is intriguing but certainly would benefit from more clarification. 
The quality of the Hubble data at the position of this subgroup is significantly poorer than in the rest of  the cluster, making it more  to identify lensed images in this area but  better data 
to come fem the HFF should help in the near future. 

A direct comparison of the 4 solutions discussed in the previous section is shown in figure \ref{fig_Mass_contour_X}.  The agreement between our 4 cases is very clear, indicating that our solution 
is robust against the assumptions made about the fiducial galaxies or the number of iterations. The same plot shows the comparison with the smoothed X-ray data. In X-rays, several prominent point 
sources match sources in the HFF data. In figure \ref{fig_Mass_contour_X} we label some of them.  All the X-ray point sources except one, have a clear counterpart in the HFF field of view. 
The only exception is source X4 in figure \ref{fig_Mass_contour_X}. At approximately one arcsecond separation we find a small group of three very faint objects where one of them may be the 
source of the X-ray emission. Deep follow-up observations at different wavelengths at this position may reveal an interesting object. 
The source labeled X1 corresponds to the foreground galaxy at z=0.1546. Radio observations reveal that this source either hosts two very powerful radio jets 
(more likely) or it lies in front of an unusually  straight shock front in the cluster \citep{vanWeeren2009} (less likely). 
In X-rays, the foreground galaxy shows up as a powerful X-ray source, possibly hosting a 
super-massive black hole at its center, a picture that is in agreement with the observed flattening of the light profile in the centre. 
This flattening is expected when a super massive black hole is at the centre of a galaxy \citep{Postman2012,Rusli2013,Thomas2014,Lopez2014}. 
Sources X2 and X3 are interesting from the lens model point of view. If we assume that X2 is at redshift z=5, our model predicts that we should expect a 
counter-image within a few arcseconds of source X3. In the optical, both X2 and X3 share a rather similar morphology with a nucleus surrounded by dusty arms. 
However in terms of galaxy size, lens magnification and relative orientation, the hypothesis that galaxies X2 and X3 are in fact the same object, loses support.  
Also, in terms of color galaxy X2 appears significantly redder than galaxy X3. 

\begin{figure}  
   \includegraphics[width=8cm]{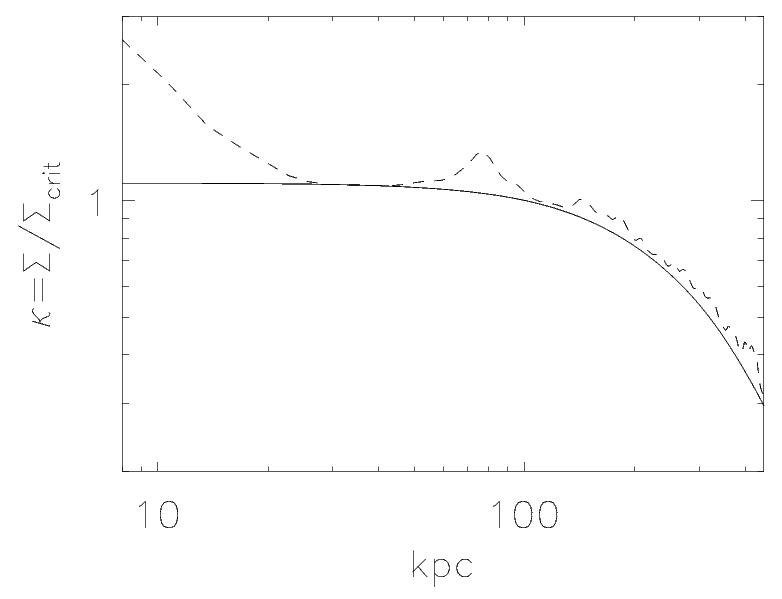}
   \caption{Profile of the solution for case IV. The mass is presented in terms of the critical surface mass density, computed for z=3 to match the mean background redshift.   The dashed line shows the total mass of galaxies plus the free-form grid and the solid line is for the grid alone corresponding to the "soft" cluster-wide mass distribution. The profile is centered in the massive galaxy member that is closer to the centre of peak C in figure \ref{fig_Deconvolved} 
below.}
    \label{fig_profile}  
\end{figure}

The profile of the mass model is presented in figure \ref{fig_profile}. Due to the lack of symmetry of this cluster, the definition, and interpretation 
of the profile is more challenging than in more relaxed and rounded clusters. In an attempt to capture some of the symmetry of the central DM peak, 
the centre of the profile in figure \ref{fig_profile} is defined at the position of the most massive galaxy that is closest to centre of the most massive peak in our  
reconstructed mass. 
This corresponds to group C (see figure \ref{fig_Deconvolved} below) as defined by \cite{Ma2009} 
(and again in \cite{Limousin2012}). More specifically, the centre is taken at RA=07:17:35.534, DEC=+37:45:0515 (J2000). 
As found in \cite{Ma2009}, and later in \cite{Limousin2012}, we find that group C is also the most massive. 
We show the profile of the total mass (dashed) and the grid component after subtraction of the contribution from the galaxies.  
The profile is strikingly shallow up to a 100 kpc, confirming earlier findings based on parametric methods \cite{Zitrin2009,Limousin2012,Medezinski2013}. 
In previous analyses based on data of two other HFF clusters, A2744 and MACS0416, we found also very shallow profiles in the central region \citep{Lam2014,Diego2014b} 
in agreement with results from other authors. The fact that these shallow profiles seem to appear in merging clusters may point in the direction that the shallowness is a 
consequence of the superposition of smaller halos in the central region. This, however, does not explain why there are not visible cusps associated to the individual halos, and 
in particular in the dominant halo in group C. Cusps are expected to survive after a cluster merger. 
In  \citep{Diego2014b} we discuss some additional possible explanations, these include mechanisms related to the baryonic component like feedback or scouring by supermassive black 
holes that are predicted to flatten the very central part of the galaxies they host. These mechanisms are however unlikely to have a significant impact on scales of tens of kpc. 
An interesting alternative discussed in the context of another HFF cluster in \citep{Diego2014b} is the possibility that dark matter has a small, but not negligible, probability 
of interaction. Simulations have shown that this mechanism, if present, is able to flatten the cusps of 
cluster halos and on the necessary scales up to 100 kpc \citep{Rocha2013}. In addition to the flattening, if DM interacts, it should exhibit a friction effect that could in 
principle be studied by possible shifts between the peaks of the DM distribution and the galaxies. To explore in this direction further it is necessary to increase the number of multiple 
images in the central regions of the cluster in order to constrain better the mass distribution. The future HFF data in this cluster may reveal enough strong lensing information to address the questions arising from our comparison more definitively.

In terms of integrated total mass of our models, we find $M_{tot}=1.27(\pm 0.21)\times10^{15} M_{\odot}$. However, we should not that our model is largely insensitive to the mass beyond the 
lensing constraints where our lens model is known to be biased towards lower masses (see \cite{Sendra2014} and references therein) so this mass should be interepreted as the integrated mass 
within the region defined by the lensing constraints. More meaningful is the total mass associated to the compact galaxy component. We find $M_{gal}=1.1(\pm 0.17)\times10^{14} M_{\odot}$. In both 
cases, the errors correspond to the dispersion of the models I,II,III and IV described in the previous section. 

\section{Variability of the solution}\label{sect_S6}

\begin{figure}  
\centerline{ \includegraphics[width=7cm]{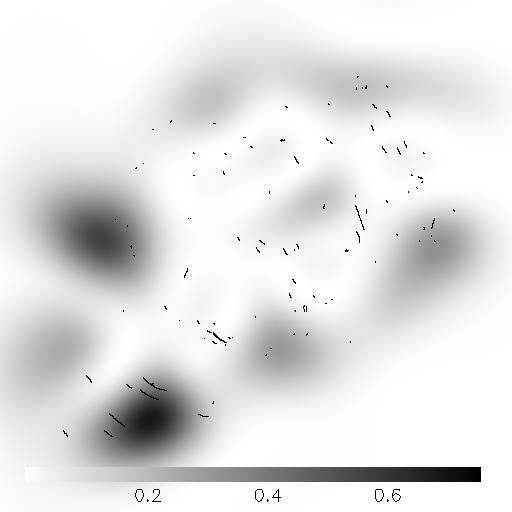}}  
   \caption{Normalized dispersion of the solutions (in percent). 
            The full data set used to do the reconstruction is also shown. 
            The larger variations in the solution concentrate around the areas where the lensing constraints start to disappear 
            or where the single arclets start to become important.} 
   \label{fig_SigmaKappa}  
\end{figure}

The motivation to produce the different solutions in cases I, II, III and IV is to study the variability of the solution with respect to our
input galaxies. As discussed in earlier works, the solution is not unique and an infinite number of models are equally consistent with the data. 
Typically, the more constraints used in the reconstruction the more similar all these infinite models are. 
In figure \ref{fig_Mass_contour_X}, a preliminary comparison of the variability of the solution was presented. The four solutions 
agree very well in general but also with some small differences that we investigate in more detail in this section.   
In figure \ref{fig_SigmaKappa} we show the dispersion of the 4 solutions normalized to the mean of the solutions and multiplied by a factor 
100 to represent percentages. The largest variation in the solutions occurs in the south-east sector of the cluster. By looking at the 
individual differences\footnote{the individual differences can be found in http://www.ifca.unican.es/users/jdiego/MACS0717} 
we check that the variability in this part of the lens is larger when comparing cases I-II, and I-III (see figure \ref{fig_diffs} in the appendix). 
That is, this variability is mostly due to the increase in the number of iterations. This is not surprising since in this region we find the 
largest concentration of individual arclets (without counterparts). These constraints, although useful, are weaker than the constraints coming 
from multiply lensed galaxies. Hence, they become more relevant only at the end of the minimization process.  In general, the largest differences 
concentrate in the region where the lensing constraints disappear.

\section{Discussion}\label{sect_S7}

The galaxy cluster MACS0717 is one of the most interesting cases for examining the interplay between baryonic and dark matter. Multiwavelength observations of this cluster have revealed powerful radio haloes that trace shock regions. The shocks are probably produced by the near simultaneous merging of several large clusters. This merging geometry  can, in principle, be understood  better through observations of the kinetic Sunyaev-Zeldovich effect of the gas in the cluster as this is proportional to the velocity along the line of sight \citep{Sayers2013,Mroczkowski2012}. Analysis of X-ray data confirms the extreme nature of this cluster with temperatures in excess of  20 keV, making this cluster one of the hottest known so far. In terms of mass, this cluster lies in the top end of what is expected for standard models with its mass distribution extending well beyond the field of view covered by the HFF observation. Weak lensing analyses \citep{Jauzac2012,Medezinski2013} reveal a filamentary structure towards the south-east from the centre and traced by the elongated galaxy distribution  \cite{Ebeling2004}. The point where the filament meets the cluster can be seen in our lens model which shows an elongation iin the south-east part of the cluster and responsible for the highly elongated lensed images we
have used in our model.

\begin{figure}  
\centerline{ \includegraphics[width=7cm]{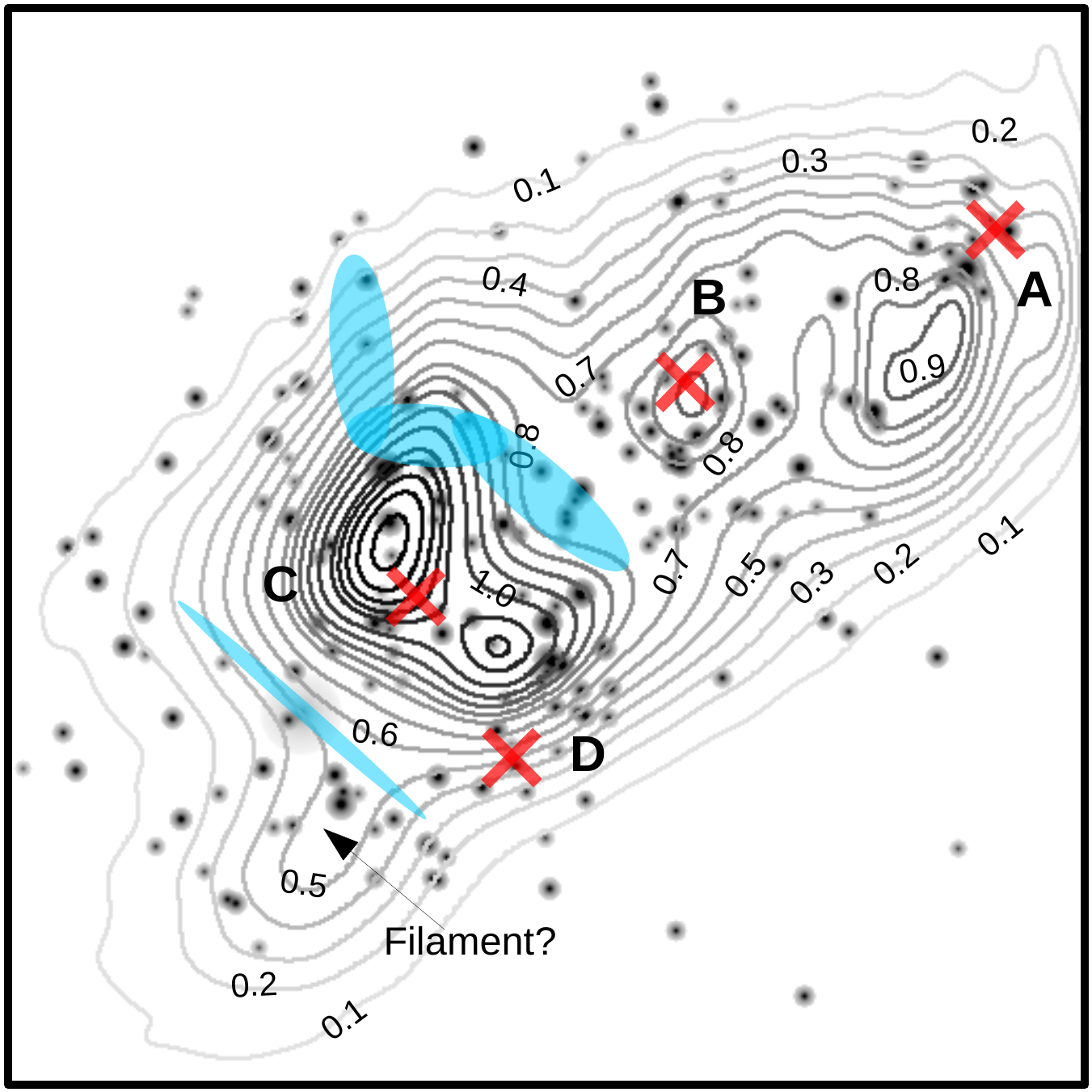}}  
   \caption{Higher resolution version of the soft, free-form cluster component of the mass map 
            obtained after a partial deconvolution. Those galaxies included  in our 
            model as small perturbations, are also shown for comparison. The contour levels are for $\kappa$ at z=3. The contour levels 
            between $\kappa=0.8$ and $\kappa=1.05$ increase in intervals of $\Delta\kappa=0.05$. The contour levels 
            above $\kappa=1.05$ increase in intervals of $\Delta\kappa=0.025$. The highest level corresponds to $\kappa=1.125$.
            The X symbols mark the position of the 4 X-ray peaks. The labels A,B,C,D denote the four subgroups seen in the optical 
            (Ma et al. 2009). The blue semitransparent regions mark the approximate position and morphology of the radio emission 
            (van Weeren et al. 2009).} 
   \label{fig_Deconvolved}  
\end{figure}

Previous lensing work has been based on parametric methods \citep{Limousin2012,Richard2014}, where the central region of the cluster has been divided into 4 separate sub-structures , relying on the optical  galaxy distribution. Interestingly the soft component of our mass model, where we have removed the  galaxy mass contribution from the total mass also shows a distribution for which 4 density maxima may be reasonably defined (figures \ref{fig_Mass_contour}, \ref{fig_Mass_contour2}, and \ref{fig_Mass_contour_X}. This sub structure is made more evident when we deconvolve partially the smooth component of our mass model. Since the smooth component is a superposition of a grid of Gaussians with a given  "pixel" scale, the resolution of the soft component is limited by 
the width chosen for  the Gaussian. Ideally this grid should be fine enough to follow all relevant substructures, but a partial deconvolution highlights the internal structure of the mass distribution after deconvolution may be even more representative of the true underlying mass distribution. Figure \ref{fig_Deconvolved}  shows the partially deconvolved version of the soft component of the mass distribution for case IV discussed above. When the deconvolved  mass distribution is compared with the  brightest galaxies from the cluster  four components of the mass distribution are now more evident. The correspondence between the location of these galaxy peaks with the general soft mass distribution (after subtracting the member galaxies) is a striking and need not have turned out this way given the freedom of our grid based modelling. In the case of the group A, on the right hand side of the HFF field at the apex of the strongly lensed region, we find a mass peak but we see that its center does not  coincide well  with the subgroup of what appear to be member galaxies on the upper right corner of the image. As discussed earlier, this could be the result of the lack of lensing constraints beyond that subgroup so certainly the reliability of this peak should not be considered as high as in the other three. Further HFF imaging may help clarify this possible interesting discrepancy. Interestingly, this offset is also found in \cite{Limousin2012} in their blind test analysis (where the mass is not assumed to trace light a priori)
highlighting the need for models (like our free-form model) that do not necesarily trace light. 
Also in \cite{Limousin2012} they also consider a five-component model in their blind analysis that the authors claim does not improve the fit further. 
However the location of one of their peaks agrees well with the elongation in the south-east part of our model. 
We should note that according to \cite{Ebeling2004}, and later to \cite{Medezinski2013}, a massive filament 
emerges from this direction of the cluster towards the sout-east, in excellent agreement with our findings (see figure \ref{fig_Deconvolved}). 
Also it is important to note that \cite{Limousin2012} did not include the arclets in the 
south-east sector that we used in our analysis, possibly explaining why their fit did not show a significant improvement in their model. 

The combined information from radio (shocks), optical, lensing and X-rays emssion it is not clear whether this cluster is in an early phase of merging  or it has gone through several core 
passages already. The radio data shows a region with prominent emission between the B group and the two western peaks (C and D) where X-ray temperatures  are also found to be higher. 
The position and geometry of the radio feature is shown in figure \ref{fig_Deconvolved} in semitransparent light blue. This picture suggests that group B and groups (C,D) are falling towards 
each other. On the other hand, 
measurments of the radial velocity \citep{Ma2009,Mroczkowski2012} suggest a significant radial component so combining the radio relic with the velocity measurements it is more likely group B is moving 
at an angle close to, but not too much, the line of sight.  The presence of an extended radio relic in the cluster suggests the system is post first core passage.  
The foreground galaxy at z=0.1546 lies in the middle of another arc with a straight geometry. This radio emission is likely produced by the AGN at this galaxy and hence not related to the cluster 
but it could be also a shock front as suggested by \cite{vanWeeren2009}, although in the same work the authors consider this option as less likely. If this straight feature originates at the cluster 
instead of the AGN, this could be a fundamental clue to determine the dynamical state of this cluster and would add evidence towards the post-merger hypothesis.  
In X-rays, the clear offset between the X-ray and luminous (or mass) peaks in group D in figure \ref{fig_Deconvolved} points to significant dynamical friction from an ongoing merging process.
On the other hand, the excellent agreement between the X-ray and mass peaks in group B points to a relatively 
mild interaction in the past of this group with the cluster environment although a small offset is also expected if B is moving in a path close to the line of sight where projection effects 
would hide any possible offsets. 

Determining the dynamical state of the cluster is important. If the cluster is in the early merging phase, 
the interpretation that the shallow central profile may be a signature of self-interacting dark matter may be more unrealistic since this mechanism to flatten central cusps 
are more efficient when clusters collide edge-on (and after first core passage) \citep{Rocha2013}. 
Of course, this possibility can not be ruled out if this cluster has gone through multiple mergers in the past. 
A confirmation of the position of the DM peak for group A may lead to important consequences to understand the interplay between baryons and DM.

\section{Conclusions}\label{sect_S8}

We have derived a free-form solution to the latest HFF data for MACS0717 and in so doing doubled the number of multiple images known. We started by  adding those systems initially uncovered by Z09 and later 
revised and/or measured in \cite{Limousin2012,Medezinski2013,Schmidt2014,Richard2014}. We then used this model and the new deeper images from the HFF program to identify 17 new multiply lensed systems doubling the number of known systems. Our method  is general so we can make use of pixel positions along the full length of the many faint extended arcs that are visible in the new HFF data. These arcs belong to images comprising  multiple systems and also individual long images for which no counter images are expected or which remain unidentified. 
Using all this information we derive a new lens model. Our new model agrees reasonably well with previous models based on a smaller number of multiply lensed images. We confirm the existence of 4 main concentrations of  dark matter in the central part of this cluster. These clumps are still clearly seen even after the contribution from the galaxies is subtracted indicating that the member galaxies must have extended cluster haloes around them. Three of the clumps correlate well with the luminous matter, but less so with the X-ray emission and  one of the clumps seem to prefer an area where no significant galaxies (or X-rays) are observed. This, however, could be a systematic effect due to the lack of constraints 
on this part of the lens. Future data from the HFF program on this cluster will help on improving the lens models even more and will open new opportunities to understand the interplay between baryonic and dark matter in this interesting cluster. 

Perhaps the most puzzling result is the very good agreement between some of the peaks of the soft dark matter component and the peaks in the X-ray and luminous matter 
while other peaks (in particular A and to some degree D) show a significant offset. 
The highly disturbed nature of this cluster, the offsets observed in some of the groups between X-ray and mass, the elongated critical curve and the presence 
of prominent radio emission suggest a collision between the groups. 
An energetic collision between clusters is expected to produce significant offsets between the galaxies (or lensing mass maps) and the plasma, as recently revealed in 
our HFF analysis of MACS0416 where distinctive offsets indicate this cluster is observed just after first core passage. These offsets are present only in groups A and D. 
The lack of offsets in group B could be explained by projection effects if B is moving close to the line of sight. The lack of offsets in group C may be understood if C is 
the most massive group and is less affected by collisions with its larger gravitational field being able to hold on to its gas (X-rays) better that the other groups. 
The offset between X-ray and mass in group D could be understood if C is moving close to the plane of the sky and towards the centre of mass of the cluster. The offsets in 
group A are the most difficult to understand although a possible explanation could be the lack of lensing constraints that do not allow to identify the peak at the right 
position. This is not an unrealistic situation if we realize that strong lensing in this cluster has long been unrecognized in the past despite being the largest 
gravitational lens known so far \citep{Zitrin2009}. 

We also confirm the shallowness of the mass profile. The earlier analysis by Z09 initially found it was very shallow and the largest known lens to date. The shallowness may be due, at least partially, to the accumulated effect of the 4 groups merging into a single massive cluster but could also be an indication of interesting physics related to the properties of  dark matter. Detailed simulations of large mergers that incorporate a degree of viscosity in DM may be able to reproduce this degree of shallowness that is known to be present as well in the other HFF clusters analyzed so far \cite{Lam2014,Diego2014b}. New data from the HFF program on this cluster will also help improve the lens models even more, offering new opportunities to understand the interplay between baryonic and dark matter in this interesting cluster.

\section{Acknowledgments}  

This work is based on observations made with the NASA/ESA {\it Hubble Space Telescope} and operated by the Association of Universities for Research in Astronomy, Inc. 
under NASA contract NAS 5-2655. The authors would like to thank the HFF team for making the data for this work available to the community.
The scientific results reported in this article are based in part on data obtained from the Chandra Data 
Archive.\footnote{ivo://ADS/Sa.CXO\#obs/16305},\footnote{ivo://ADS/Sa.CXO\#obs/04200},\footnote{ivo://ADS/Sa.CXO\#obs/01655} 
We would like to thank Harald Ebeling for making  the code {\small ASMOOTH} \citep{Ebeling2006} available. 
T.J.B. thanks the University of Hong Kong for generous hospitality. J.M.D acknowledges support of the consolider project 
CAD2010-00064 and AYA2012-39475-C02-01 funded by the Ministerio de Economia y Competitividad. 
AZ was provided by NASA through Hubble Fellowship grant \#HST-HF2-51334.001-A awarded by STScI.
The authors thank Francisco Carrera, Nanda Rea and Marceau Limousin for very useful comments. 
  
\label{lastpage}
\bibliographystyle{mn2e}
\bibliography{MyBiblio} 

\newpage
 
\appendix

\section{Compilation of arc positions}

\begin{table*}
    \begin{minipage}{115mm}                                               
    \caption{Full lensing data set. The first column shows the system ID following the original notation 
             of Z09. 
             The second and third columns show the coordinates of each arclet. 
             Column 4 includes the redshifts used in our study (taken from Z09 and  Schmidt et al. 2014 when appropriate).
             Some of these redshifts are photometric and some are based on colour and/or predicted by the lens model. 
	     The last column contains additional useful information. 
             New images are denoted with {\it N} in the {\it Notes} column. 
             Spectroscopic redshifts are marked with an {\it S} in this column. The redshifts that are obtained from 
             Z09 are denoted {\it Z}. When the redshifts come from Schmidt et al. (2014) ,we denote it with {\it Sc}.
             The system 19 confirmed by Vanzella at al. (2014) 
             is denoted as {\it V}.}
 \label{tab1}
 \begin{tabular}{ccccc}   

  ID     &  RAJ2000(h:m:s)  & DECJ2000(d:m:s)  &    z    & Notes  \\

  1.1   &   7:17:34.865     &   +37:44:28.39   &   2.963  &  S,Z\\
  1.2   &   7:17:34.514     &   +37:44:24.43   &   2.963  &  S,Z\\
  1.3   &   7:17:33.823     &   +37:44:17.88   &   2.963  &  S,Z\\
  1.4   &   7:17:32.234     &   +37:44:13.13   &   2.963  &  S,Z\\
  1.5   &   7:17:37.384     &   +37:45:40.95   &   2.963  &  S,Z\\
  2.1   &   7:17:34.267     &   +37:44:27.72   &   2.500  &     \\
  2.2   &   7:17:33.691     &   +37:44:21.30   &   2.500  &     \\
  3.1   &   7:17:35.645     &   +37:44:29.44   &   1.855  &  S,Sc\\
  3.2   &   7:17:34.656     &   +37:44:21.08   &   1.855  &  S,Sc\\
  3.3   &   7:17:37.702     &   +37:45:13.86   &   1.855  &  S,Sc\\
  4.1   &   7:17:31.440     &   +37:45:01.55   &   1.855  &  S,Sc\\
  4.2   &   7:17:30.324     &   +37:44:40.70   &   1.855  &  S,Sc\\
  4.3   &   7:17:33.828     &   +37:45:47.77   &   1.855  &  S,Sc\\
  5.1   &   7:17:31.169     &   +37:44:48.73   &   4.300  &     \\
  5.2   &   7:17:30.694     &   +37:44:34.19   &   4.300  &     \\
  5.3   &   7:17:36.000     &   +37:46:02.75   &   4.300  &     \\
  5.4   &   7:17:32.657     &   +37:44:31.33   &   4.300  &  N  \\
  6.1   &   7:17:27.434     &   +37:45:25.56   &   2.100  &     \\
  6.2   &   7:17:27.041     &   +37:45:09.90   &   2.100  &     \\
  6.3   &   7:17:29.734     &   +37:46:11.21   &   2.100  &     \\
  7.1   &   7:17:27.970     &   +37:45:58.90   &   2.200  &     \\
  7.2   &   7:17:27.607     &   +37:45:50.87   &   2.200  &     \\
  7.3   &   7:17:26.160     &   +37:45:06.59   &   2.200  &     \\
  8.1   &   7:17:27.982     &   +37:46:10.81   &   2.300  &     \\
  8.2   &   7:17:26.890     &   +37:45:47.41   &   2.300  &     \\
  8.3   &   7:17:25.566     &   +37:45:06.88   &   2.300  &     \\
  12.1  &   7:17:32.438     &   +37:45:06.80   &   1.699  &  S,Sc\\
  12.2  &   7:17:30.617     &   +37:44:34.51   &   1.699  &  S,Sc\\
  12.3  &   7:17:33.890     &   +37:45:38.38   &   1.699  &  S,Sc\\
  13.1  &   7:17:32.522     &   +37:45:02.30   &   2.547  &  S,Z\\
  13.2  &   7:17:30.610     &   +37:44:22.85   &   2.547  &  S,Z\\
  13.3  &   7:17:35.083     &   +37:45:48.20   &   2.547  &  S,Z\\
  14.1  &   7:17:33.305     &   +37:45:07.96   &   1.855  &  S,Sc\\
  14.2  &   7:17:31.111     &   +37:44:22.92   &   1.855  &  S,Sc\\
  14.3  &   7:17:35.076     &   +37:45:37.19   &   1.855  &  S,Sc\\
  15.1  &   7:17:28.253     &   +37:46:19.24   &   2.405  &  S,Z\\
  15.2  &   7:17:26.090     &   +37:45:36.29   &   2.405  &  S,Z\\
  15.3  &   7:17:25.584     &   +37:45:16.20   &   2.405  &  S,Z\\
  16.1  &   7:17:28.589     &   +37:46:23.88   &   3.100  &     \\
  16.2  &   7:17:26.050     &   +37:45:34.49   &   3.100  &     \\
  16.3  &   7:17:25.661     &   +37:45:13.43   &   3.100  &     \\
  17.1  &   7:17:28.646     &   +37:46:18.55   &   2.500  &     \\
  17.2  &   7:17:26.256     &   +37:45:31.82   &   2.500  &     \\
  17.3  &   7:17:25.966     &   +37:45:12.71   &   2.500  &     \\
  17.4  &   7:17:26.592     &   +37:45:29.84   &   2.500  &  N  \\
  18.1  &   7:17:27.406     &   +37:46:07.10   &   2.400  &     \\
  18.2  &   7:17:26.683     &   +37:45:51.66   &   2.400  &     \\
  19.1  &   7:17:38.170     &   +37:45:16.87   &   6.387    &  S,V\\ 
  19.2  &   7:17:37.860     &   +37:44:33.87   &   6.387    &  S,V\\   

\end{tabular}
\end{minipage}
\end{table*}

\setcounter{table}{0}
\begin{table*}
    \begin{minipage}{115mm}                                               
    \caption{cont.}

 \begin{tabular}{ccccc}    

  ID     &  RAJ2000(h:m:s)  & DECJ2000(d:m:s)  &    z    & Notes  \\

  20.1  &   7:17:29.804     &   +37:45:54.42   &   5.000  &  N  \\
  20.2  &   7:17:29.612     &   +37:45:52.62   &   5.000  &  N  \\
  21.1  &   7:17:28.524     &   +37:45:17.94   &   2.000  &  N  \\
  22.1  &   7:17:28.256     &   +37:45:21.23   &   1.500  &  N  \\
  22.2  &   7:17:28.236     &   +37:45:19.91   &   1.500  &  N  \\
  23.1  &   7:17:31.107     &   +37:45:46.47   &   3.000  &  N  \\
  23.2  &   7:17:30.953     &   +37:45:43.14   &   3.000  &  N  \\
  24.1  &   7:17:28.565     &   +37:45:08.76   &   3.000  &  N  \\
  25.1  &   7:17:31.269     &   +37:44:41.10   &   4.500  &  N  \\
  25.2  &   7:17:31.083     &   +37:44:33.92   &   4.500  &  N  \\
  25.3  &   7:17:36.699     &   +37:45:59.07   &   4.500  &  N  \\
  27.1  &   7:17:35.369     &   +37:44:52.50   &   2.000  &  N  \\
  27.2  &   7:17:35.413     &   +37:44:51.16   &   2.000  &  N  \\
  28.1  &   7:17:37.675     &   +37:43:58.70   &   2.000  &  N  \\
  29.1  &   7:17:36.211     &   +37:44:35.43   &   1.800  &  N  \\
  29.2  &   7:17:34.290     &   +37:44:18.97   &   1.800  &  N  \\
  29.3  &   7:17:37.461     &   +37:44:59.83   &   1.800  &  N  \\
  30.1  &   7:17:34.055     &   +37:44:20.78   &   1.800  &  N  \\
  30.2  &   7:17:37.569     &   +37:45:03.96   &   1.800  &  N  \\
  31.1  &   7:17:29.928     &   +37:45:22.68   &   1.750  &  N  \\
  31.2  &   7:17:29.043     &   +37:45:01.98   &   1.750  &  N  \\
  31.3  &   7:17:31.588     &   +37:45:53.94   &   1.750  &  N  \\
  32.1  &   7:17:28.671     &   +37:45:27.87   &   2.700  &  N  \\
  32.2  &   7:17:31.426     &   +37:46:09.57   &   2.700  &  N  \\
  32.3  &   7:17:27.895     &   +37:44:56.93   &   2.700  &  N  \\
  33.1  &   7:17:32.102     &   +37:45:29.53   &   5.000  &  N  \\
  33.2  &   7:17:32.777     &   +37:45:50.83   &   5.000  &  N  \\
  33.3  &   7:17:28.884     &   +37:44:19.29   &   5.000  &  N  \\
  34.1  &   7:17:33.090     &   +37:45:55.36   &   2.300  &  N  \\
  34.2  &   7:17:30.979     &   +37:45:04.33   &   2.300  &  N  \\
  34.3  &   7:17:29.592     &   +37:44:39.16   &   2.300  &  N  \\
  36.1  &   7:17:27.446     &   +37:46:19.25   &   2.500  &  N  \\
  36.2  &   7:17:25.974     &   +37:45:47.89   &   2.500  &  N  \\
  36.3  &   7:17:24.797     &   +37:45:21.06   &   2.500  &  N  \\
  37.1  &   7:17:35.305     &   +37:45:17.08   &   4.000  &  N  \\
  37.2  &   7:17:35.203     &   +37:45:17.08   &   4.000  &  N  \\
  39.1  &   7:17:36.881     &   +37:43:54.51   &   2.000  &  N  \\
  40.1  &   7:17:40.154     &   +37:43:36.19   &   3.000  &  N  \\
  41.1  &   7:17:37.978     &   +37:43:41.06   &   3.000  &  N  \\
  42.1  &   7:17:39.220     &   +37:44:01.71   &   3.000  &  N  \\
  43.1  &   7:17:34.668     &   +37:43:44.25   &   3.000  &  N  \\
  44.1  &   7:17:36.658     &   +37:43:58.23   &   2.000  &  N  \\
  45.1  &   7:17:33.556     &   +37:44:21.12   &   3.000  &  N  \\
  45.2  &   7:17:32.029     &   +37:44:16.37   &   3.000  &  N  \\
  45.3  &   7:17:37.071     &   +37:45:43.00   &   3.000  &  N  \\
  47.1  &   7:17:38.397     &   +37:43:35.46   &   3.000  &  N  \\
  49.1  &   7:17:36.709     &   +37:43:59.64   &   3.000  &  N  \\
  49.2  &   7:17:34.316     &   +37:43:50.64   &   3.000  &  N  \\
  50.1  &   7:17:29.866     &   +37:44:37.45   &   3.000  &  N  \\
  50.2  &   7:17:31.100     &   +37:45:02.55   &   3.000  &  N  \\
  50.3  &   7:17:34.273     &   +37:46:01.85   &   3.000  &  N  \\
  52.1  &   7:17:28.407     &   +37:46:18.27   &   3.000  &  N  \\
  52.2  &   7:17:26.455     &   +37:45:37.79   &   3.000  &  N  \\
  52.3  &   7:17:25.689     &   +37:45:08.95   &   3.000  &  N  \\

 \end{tabular}
 \end{minipage}
\end{table*}

\begin{table*}
    \begin{minipage}{115mm}                                               
    \caption{Additional candidate systems not used in the main analysis of this work. 
             At the time of submission of this paper, additional observations in the optical bands 
             have been made available that has allowed us to uncover new additional candidate systems. All these systems 
             are in excellent agreement with the model presented in this paper so their inclusion in the analysis should result 
             in small changes to the mass model. Stamps of these systems are also provided in the website (http://www.ifca.unican.es/users/jdiego/MACS0717). 
             These systems, together with the new ones that may be identified after the impending release of the new IR data, will be used in a subsequent paper.}
 \label{tab2}
 \begin{tabular}{cccc}   

  ID     &  RAJ2000(h:m:s)  & DECJ2000(d:m:s)  &    z    \\

 53.1   &   7:17:28.843     &   +37:45:40.60   &   2.7  \\
 53.2   &   7:17:31.035     &   +37:46:05.71   &   2.7  \\
 53.3   &   7:17:27.619     &   +37:44:49.79   &   2.7  \\
 54.1   &   7:17:36.038     &   +37:44:42.81   &   1.2  \\
 54.2   &   7:17:35.881     &   +37:44:40.50   &   1.2  \\
 55.1   &   7:17:37.391     &   +37:43:54.66   &   5.0  \\
 55.2   &   7:17:35.290     &   +37:43:44.35   &   5.0  \\
 56.1   &   7:17:34.354     &   +37:45:46.06   &   5.0  \\
 56.2   &   7:17:32.964     &   +37:45:21.47   &   5.0  \\
 56.3   &   7:17:30.180     &   +37:44:08.24   &   5.0  \\
 57.1   &   7:17:34.318     &   +37:45:43.93   &   5.0  \\
 57.2   &   7:17:33.131     &   +37:45:23.45   &   5.0  \\
 57.3   &   7:17:30.225     &   +37:44:06.47   &   5.0  \\
 58.1   &   7:17:34.321     &   +37:45:44.60   &   5.0  \\
 58.2   &   7:17:33.078     &   +37:45:23.01   &   5.0  \\
 58.3   &   7:17:30.197     &   +37:44:06.64   &   5.0  \\
 59.1   &   7:17:34.271     &   +37:45:45.49   &   4.0  \\
 59.2   &   7:17:33.116     &   +37:45:24.60   &   4.0  \\
 60.1   &   7:17:36.614     &   +37:45:48.98   &   2.6  \\
 60.2   &   7:17:31.455     &   +37:44:26.40   &   2.6  \\
 60.3   &   7:17:31.351     &   +37:44:28.87   &   2.6  \\
 60.4   &   7:17:32.930     &   +37:44:28.56   &   2.6  \\
 60.5   &   7:17:31.846     &   +37:44:40.95   &   2.6  \\
 61.1   &   7:17:34.125     &   +37:45:37.75   &   2.4  \\
 61.2   &   7:17:33.448     &   +37:45:26.19   &   2.4  \\
 62.1   &   7:17:31.082     &   +37:44:42.36   &   3.0  \\
 62.2   &   7:17:30.965     &   +37:44:38.74   &   3.0  \\
 63.1   &   7:17:29.413     &   +37:44:58.33   &   3.0  \\
 63.2   &   7:17:29.588     &   +37:45:04.63   &   3.0  \\
 64.1   &   7:17:33.044     &   +37:44:19.95   &   2.5  \\
 64.2   &   7:17:32.146     &   +37:44:18.22   &   2.5  \\
 65.1   &   7:17:33.272     &   +37:44:21.32   &   4.0  \\
 65.2   &   7:17:31.682     &   +37:44:18.37   &   4.0  \\

 \end{tabular}
 \end{minipage}
\end{table*}

\begin{figure*}  
\centerline{ \includegraphics[width=14cm]{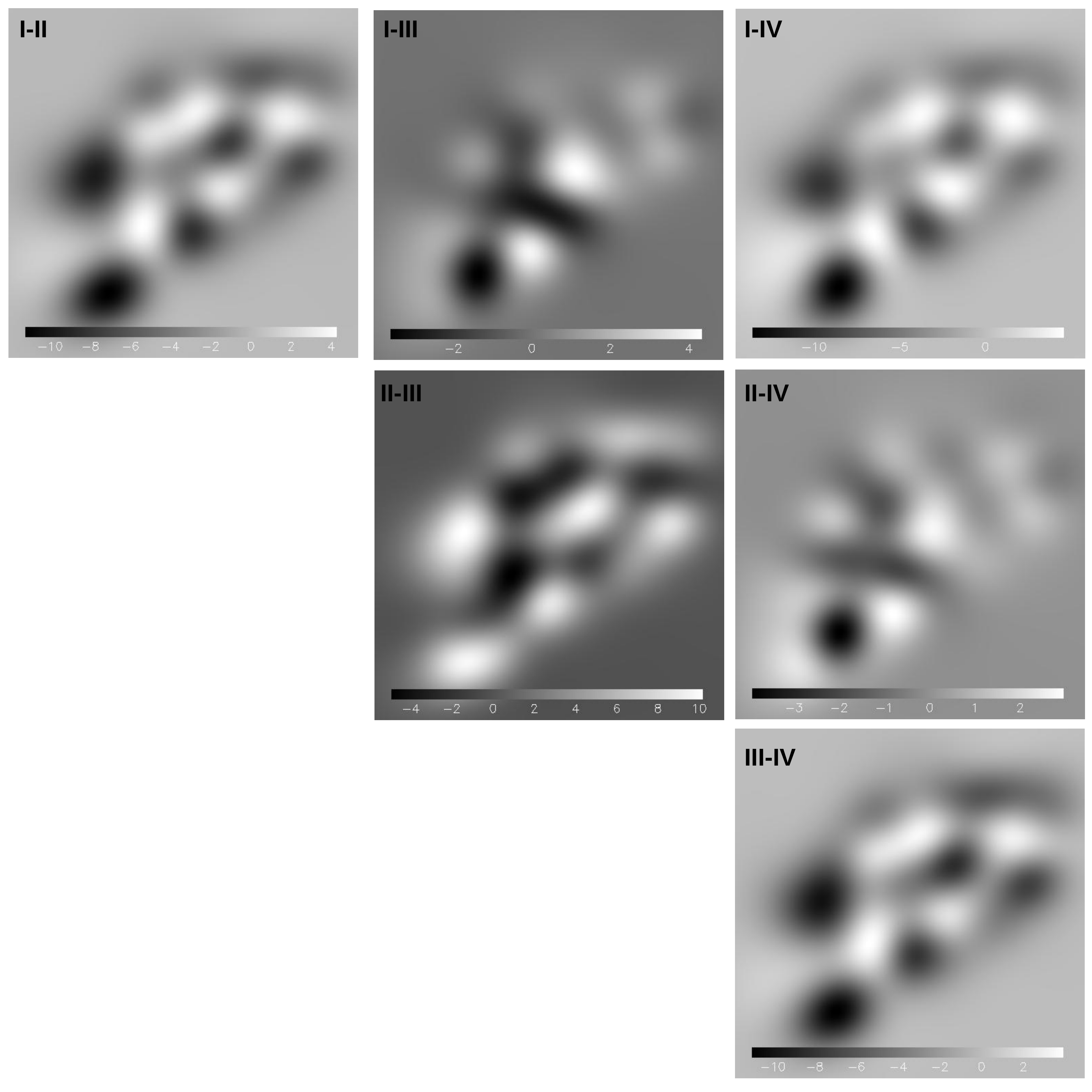}}  
   \caption{Normalized differences (in percent) between the models I,II,III, and IV.} 
   \label{fig_diffs}  
\end{figure*}

\end{document}